\documentclass[11pt]{article}

\usepackage[latin1]{inputenc}
\usepackage{amsmath}
\usepackage{amsfonts}
\usepackage{amssymb}
\usepackage{fix-cm}
\usepackage{graphicx}
\usepackage{array}
\usepackage{multirow}
\usepackage[left=1in, right=1.00in, top=1in, bottom=1.0in]{geometry}
\usepackage{url}
\usepackage{color}
\usepackage{setspace}
\usepackage{wrapfig}
\usepackage[scriptsize]{caption}
\usepackage{soul}
\usepackage[small]{titlesec}
\usepackage{enumerate}
\usepackage[shortlabels]{enumitem}
\usepackage{float}
\usepackage{comment}
\usepackage{booktabs}

\date{}

\title{Prevalence of Upper Extremity Distal Predominant Weakness Pattern in Chronic Stroke}
\author{Ryan H. Baxter$^{1}$, Kotaro Tsutsumi$^{1}$, Marc W. Slutzky$^{2}$, An H. Do$^{1}$,}

\begin{document}

\maketitle

$^{1}$University of California (UCI), Department of Neurology, Irvine, CA, USA\\ 

$^{2}$ Departments of Neurology, Neuroscience, Physical Medicine and Rehabilitation, and Biomedical Engineering, \\Northwestern University, Chicago, IL, USA\\

Email: rhbaxter@hs.uci.edu, and@uci.edu

\section*{Abstract}

\textbf{Background and Objectives:} Hemiparesis following subcortical stroke has been classically thought to be greater in the distal upper extremity (UE) compared to proximal UE.  
However, there is a lack of extensive prevalence data for this distal predominant weakness in chronic stroke. To address this knowledge gap, a retrospective study was undertaken. This study seeks to examine the prevalence of the distal predominant weakness pattern among patients with exclusively subcortical chronic stroke compared to those with other stroke distributions, investigate additional differences between these patient cohorts, and use the study's database to draw broader conclusions regarding chronic strokes as a whole.

\textbf{Methods:} Specifically, outpatient medical records at the University of California, Irvine were retrospectively reviewed to identify chronic stroke subjects. Identified subjects were classified based on their stroke location as either exclusively subcortical or not, based on radiographic data. In addition, subjects were also classified as either exhibiting a distal predominant weakness pattern or not.
The prevalence of distal predominant weakness was compared between subcortical and non-subcortical locations using a $\chi$-squared test.

\textbf{Results:} A total of 250 chronic stroke subjects (mean time post-stroke: 861 days) were enrolled into this study. 
When a whole-brain definition of exclusively subcortical stroke was applied, 30.6\% of subjects with exclusively subcortical stroke compared to 17.4\% of non-exclusively subcortical stroke subjects exhibited a distal predominant weakness (OR = 2.09, 95\% CI 1.15-3.81, p=0.014).
When a supratentorial definition of exclusively subcortical stroke was applied (i.e., brainstem lesions excluded), the 27.9\% of subjects with exclusively subcortical stroke compared to 17.9\% of non-exclusively subcortical stroke subjects exhibited a distal predominant weakness (OR = 2.16, 95\% CI 1.17-3.96, p=0.012).

\textbf{Discussion:} These findings suggest that the distal predominant UE weakness is more prevalent among chronic subcortical compared to non-subcortical stroke. 
Furthermore, although most chronic stroke subjects recovered without UE weakness (60\%), the distal predominant weakness was the most common UE weakness pattern observed in this cohort of chronic stroke subjects (23\%), followed by a uniform weakness throughout the UE (12\%). Conversely, a proximal predominant weakness pattern was rare (3\%).
These findings provide prevalence data regarding the distal UE weakness patterns. Such information can predict long-term outcomes based on stroke location and aid in the planning of rehabilitation care. It may also aid physicians in stroke lesion localization and guide where the public health emphasis should be placed in stroke rehabilitation research. 

\section{Introduction}

Chronic stroke patients are often affected by permanent motor deficits. Patterns of weakness have been noted among specific cohorts of stroke patients, such as distal predominant upper extremity weakness patterns \cite{lindj:23,Beebe:08,Schneider:94,Tyson:05}.  
Classically, it is generally believed that strokes in subcortical regions are more likely to leave the patient with a distal predominant weakness pattern in the upper extremities.  Current teaching seems to accept this without explicit epidemiological data in support of this claim \cite{Castle-Kirszbaum:20}, \cite{Kandel:91}, \cite{Merholz:19}.
Such chronic stroke patients have weakness in the distal muscles (i.e., hand and wrist muscles) more than the proximal ones (shoulder and elbow flexors/extensors). Meanwhile, strokes not exclusively located in these areas are thought to affect the limb more uniformly \cite{Kandel:91}.
Surprisingly, the prevalence of such upper extremity weakness patterns in the context of stroke location has not been explicitly characterized  for chronic stroke survivors. 
Previous prospective studies have only examined the subacute stroke populations \cite{lindj:23}, and no large-scale study to our knowledge has examined this phenomenon in chronic stroke patients

Therefore, the aims of this study were to examine the prevalence of the distal predominant weakness pattern among patients with exclusively subcortical chronic stroke compared to those with other stroke distributions, understand the prevalence of other weakness patterns after stroke, and investigate the role of stroke type in weakness patterns.

\section{Methods}
\subsection{Study Design Overview}
 This study is a retrospective study on the prevalence of distal predominant weakness in chronic stroke patients. A chart review was conducted to identify suitable chronic stroke patients who have sufficient strength data available (criteria defined below) and have no coexisting conditions that can cause neurological impairment, as defined further below. Study candidates were adults diagnosed with chronic hemorrhagic or ischemic strokes and without other diseases or conditions that might cause neurological deficits. From the selected charts, stroke location, upper extremity strength values, and patient demographic information were recorded. Data were analyzed to investigate the prevalence of distal predominant weakness in addition to secondary analyses. 

\vspace{-0.075in}

\subsection{Subjects and Chart Review}
This study was approved by the University of California, Irvine (UCI) Institutional Review Board (IRB)
and was declared exempt.
Subjects were identified for this study by reviewing outpatient records from UCI stroke,  physiotherapy, rehabilitation, and neurosurgery clinics. 
Additionally, data extraction from the UCI electronic medical record (EMR) system was requested from the UCI Health Services Office of Information Technology. This extraction included a list of all patients within the UCI EMR system with a documented stroke within 10 years of the request date. This search included the ICD9 codes 433.x1, 434.x1, and 436, 431.x and ICD10 codes 163.x, 164.x, and 161.x.

From this list, EMR charts from potential subjects were sequentially selected for manual review of inclusion/exclusion criteria. Patient data was collected until the recruitment goal of 250 enrolled subjects was reached. This represents a convenience sample size, as no comparable studies were available to properly perform an \textit{a priori} power analysis. Patients meeting the following inclusion criteria were included: 1. Age $\ge$ 18 years old at time of stroke; 2. Patients with evaluations $\ge$ 180 days post-stroke; 3. Manual muscle testing data available at $\ge$ 180 days post-stroke.

Exclusion criteria were as follows: 1. Age $<$18 years at date of stroke; 2. Coexisting conditions that can create neurologic or musculoskeletal weakness, such as brain tumors, demyelinating syndromes, amputations to upper extremities, neuropathy, etc.; 3.  Diagnosis of subdural hematoma, sub-arachnoid hemorrhage, or transient ischemic attack.

Subjects who met these criteria were subsequently selected for the study, and their charts were used to extract data related to upper extremity muscle strength and stroke location (described in detail further below). In addition, pertinent patient demographic information (age, sex, race, and ethnicity), stroke type, stroke location, and upper extremity muscle strength evaluations were also extracted and recorded. 

\vspace{-0.075in}
\subsection{Stroke Location}
\vspace{-0.05in}
The subjects' stroke location was determined based on the radiologist's evaluation of each subjects' MRI or CT head scan. This information was identified directly from the radiologists' impression note completed at the time of stroke workup. Potential subjects must have at least one available appropriately timed image available. If multiple scans were available, the most recent evaluation subsequent to the most recent stroke was used. If the radiologist's evaluation provided insufficient information for our purposes, we manually examined MRI and CT scans in order to determine the stroke's topography.
Based on these steps, we classified each stroke into one or more of the following categories: cortex, corona radiata, deep structures (including the basal ganglia, thalamus, and internal capsule), brainstem, and cerebellum.

\subsubsection{Defining Exclusively Subcortical and Non-exclusively Subcortical Strokes}

Given the variable definitions of subcortical brain regions (\cite{Chumin:22}), two definitions of exclusively subcortical stroke locations were adopted as follows. First, a ''whole-brain'' definition of subcortical strokes was defined as stroke being located in the deep structures, corona radiata, brainstem, or more than one of these locations. Second, a supratentorial definition of subcortical strokes was defined as those occurring in the deep structures, corona radiata, or more than one of these locations (excluding brainstem). 
Non-exclusively, subcortical strokes were defined as affecting areas outside these regions, regardless of any subcortical involvement. Each of these two classifications of exclusively subcortical strokes were later individually compared to non-exclusively subcortical strokes to examine the prevalence of distal predominant weakness patterns in each population.

\subsection{Upper Extremity Muscle Strength Evaluations}

To characterize the pattern of subjects' weakness, we extracted health care providers' physical examination of subjects' muscle strength on the stroke affected side using the Medical Research Council Manual Muscle Testing (MMT) scale (0 = no contractions felt in the muscle, 1 = trace contractions, 2 = active movement with gravity eliminated, 3 = active movement against gravity, 4 = active movement against gravity with some resistance, 5 = active movement against full resistance). Specifically, shoulder (forward) flexion and extension, elbow flexion and extension, wrist flexion and extension, and finger flexion and extension were evaluated (total of 8 MMT values expected). Where applicable, names of muscles were accepted as synonyms for the joints above (e.g., shoulder and deltoid, elbow and bicep/triceps, etc.).

To compare proximal and distal muscle groups, we first defined a patient's "weakness score" mathematically as:

\textit{Weakness Score = Mean proximal strength - mean distal strength}

where shoulder flexors/extensors and elbow flexors/extensors are proximal muscle groups while wrist flexors/extensors and finger flexors/extensors are  distal muscle groups.

For primary analysis (details in Section \ref{sec:primary}), two classes of subjects were defined:
\begin{itemize}
    \item Subjects with weakness scores$>$0, or with qualitative data (details further below) describing the subject's upper extremity as having distal predominant weakness, were defined as exhibiting a distal predominant weakness pattern.
    \item Subjects with weakness scores$\leq$0, or with qualitative data describing the subject's upper extremity as not having distal predominant weakness, were defined as not exhibiting a distal predominant weakness pattern.
\end{itemize}

We defined the following classes for secondary analyses (details in Section \ref{sec:secondary}):
\begin{itemize}
    \item Subjects with weakness scores$<$0 would be classified as displaying proximal predominant weakness; i.e. their proximal muscle groups were found to be weaker than their distal muscle groups.
    \item Subjects with weakness scores=0, and exhibited some weakness in any muscle, would be classified as having equal weakness throughout the limb.
    \item Subjects with weakness scores=0, and exhibited no weakness in any muscle, would be classified as having no weakness.
    \item Subjects with weakness scores$>$0 would be classified as displaying distal predominant weakness; i.e. their distal muscle groups were found to be weaker than their proximal muscle groups.
\end{itemize}

If MMT data were incomplete for a given potential subject, the following approach was used:

\textbf{A.} If only a single strength value was available for the entire upper limb, the potential subject was automatically excluded.

\textbf{B.} If there were more than two strength values missing per limb segment (distal groups or proximal groups), the potential subject was automatically excluded. If the missing data points could be found from another note (with preference for the most recent eligible note), these could be collected and substituted for the missing values, and the potential subject could be included. 

\textbf{C.} Since providers frequently document only one strength value for the shoulder (deltoids), a single strength value was allowed to be used to describe both shoulder flexion and extension.

\textbf{D.} If quantitative strength records were missing but there were qualitative documentation regarding the distal and proximal strength relationship present, we automatically classified this subject according to the description (i.e., distal predominant weakness pattern or no distal predominant weakness pattern). 

This data was extracted from provider notes using the following preference hierarchy, due to each specialty's focus on examination : 1. Physiatry physicians; 2. physical/occupational therapists; 3. neurologists; 4. neurosurgeons. 
Only notes that were taken at least six months after the date of the stroke were considered. If multiple notes were found in this time frame, the most recent note at the time of data collection was considered.

\subsection{Primary Analysis}\label{sec:primary}

Subjects were then categorized into exclusively subcortical and non-exclusively subcortical strokes groups using both the whole brain and supratentorial definitions.
To determine whether there was a significant difference between weakness patterns in subcortical and not-exclusively-subcortical strokes, a $\chi$-squared analysis was completed.  All significance was evaluated using a threshold of \(\alpha=0.05\). An odds ratio (OR) was calculated between the exclusively subcortical and non-exclusively subcortical populations.

\subsection{Secondary Analyses}\label{sec:secondary}

Data were also analyzed to examine whether stroke etiology played a role in chronic stroke weakness patterns, particularly, given that hemorrhagic strokes are more commonly subcortical \cite{Chen:14,Rymer:11,Ojaghihaghighi:17,Magid-Bernstein:22}. To this end, all cases were classified into ischemic or hemorrhagic based on radiological imaging. Using previously identified weakness patterns, a $\chi$-squared test and OR were used to characterize whether stroke etiology affected the weakness pattern.
To determine if stroke location was affected by stroke etiology, a $\chi$-squared analysis was completed.

\section{Results}
\subsection{Study Subjects Demographic Information}

\begin{figure}[htpb!]
    \centering
    \includegraphics[width=0.75\linewidth]{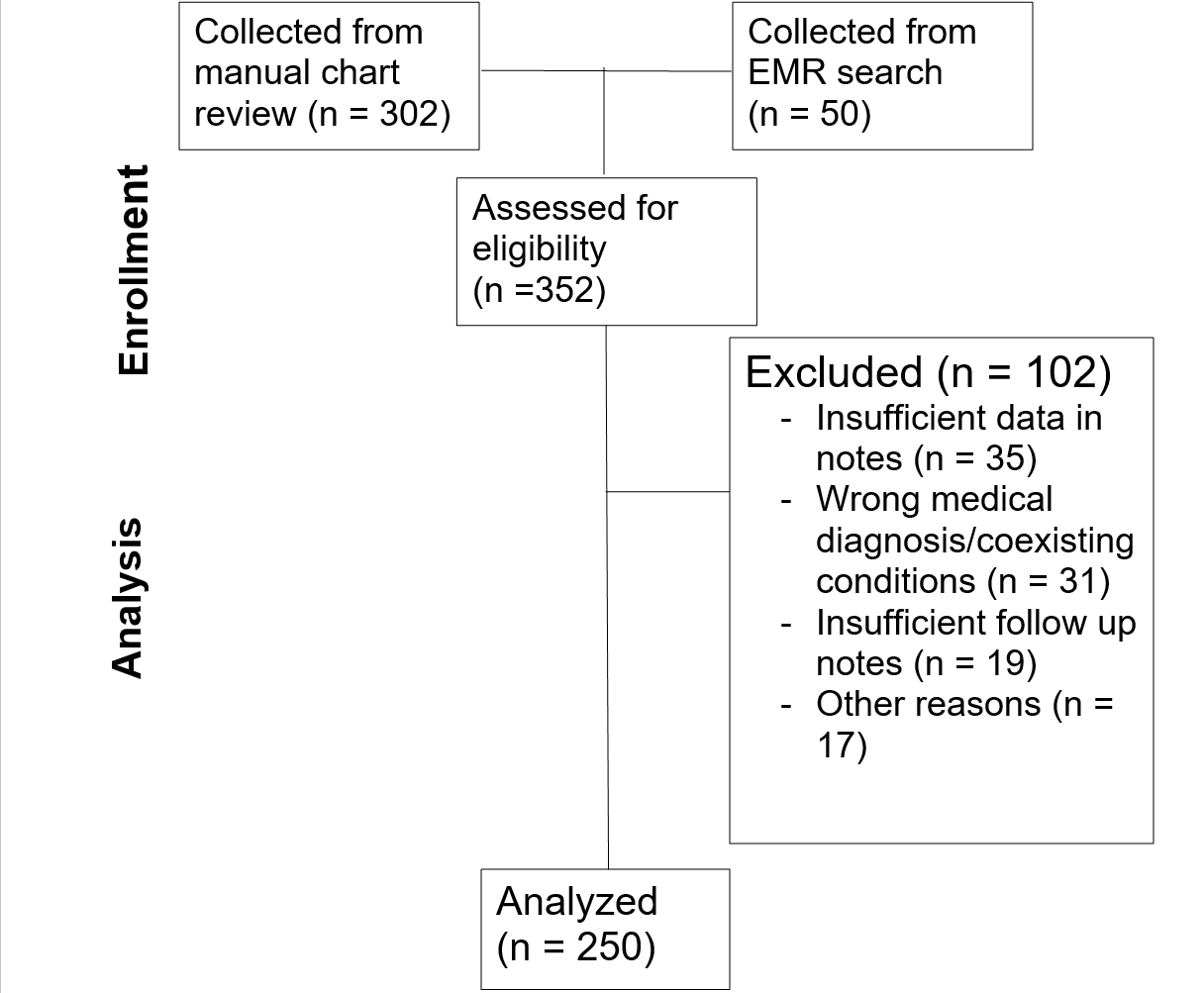}
    \caption{CONSORT Diagram Describing Inclusion}
    \label{fig:consortdiagram}
\end{figure}

A total of 352 potential subjects, who were seen at UC Irvine outpatient facilities were evaluated retrospectively  (Fig. \ref{fig:consortdiagram}). A total of 250 subjects were included and the rest were excluded with the leading cause being insufficient data in notes and coexisting conditions affecting the patient's neurological status. Of the 250 subjects included, 15 had  qualitative data only, as described in the methods section. Of these, 10 subjects were described as having distal predominant weakness, 1 displayed no weakness, and 4 were only described as not showing a distal predominant weakness pattern.

\subsection{Subject demographics}

Subject demographics are summarized in Table \ref{tab:demographics}. No statistically significant differences in sex, age, race, or ethnicity between any groups were found (p$>$0.05 for all categories via a $\chi$-squared analysis). We identified that 40.4\% of the cohort had exclusively subcortical strokes based on the supratentorial definition of subcortical strokes, while 48.8\% of the cohort was found to exhibit the whole-brain definition of subcortical strokes (Table \ref{tab:outcomes}). Nearly 80\% of all strokes were ischemic, a figure similar to the 87\% rate reported globally \cite{Martin:25}. The average time between the date of stroke and that of the assessment used for this study was 869 days$\pm$ 811.2 (range 185 - 11,373 days).

\begin{table*}[htpb!]
      \caption{Demographic data of the entire cohort and the exclusively subcortical population}
    \begin{center}
    \label{demographicdata}
    \scriptsize
    \begin{tabular}{|>{\centering\arraybackslash}p{0.15\linewidth}|c|>{\centering\arraybackslash}p{0.15\linewidth}|>{\centering\arraybackslash}p{0.15\linewidth}|c|} \hline  
    \centering
         &  {Entire Cohort - N=250 (\%)}& {Exclusively Subcortical - Supratentorial - N=100 (\%})&  Exclusively subcortical - Whole-brain - N=121 (\%)&{Non-exclusively subcortical - N=150 (\%})\\ \hline  
         \textbf{Sex}&  & &  &\\ \hline  
         Male&  161 (64.4)& 67 (67)&  76 (62.8)&94 (62.7)\\ \hline  
         Female&  81 (32.4)& 30 (30)&  42 (34.7)&51 (34)\\ \hline  
         Unknown/not reported&  8 (32)& 3 (3)&  3 (2.5)&4 (2.7)\\ \hline  
         \textbf{Race}&  & &  &\\ \hline  
         Asian&  67 (26.8)& 30 (30)&  36 (29.8)&37 (24.7)\\ \hline  
         Black/African-American&  5 (2)& 3 (3)&  3 (2.5)&2 (1.3)\\ \hline  
         Native Hawaiian/Pacific Islander&  1 (0.4)& 1 (1)&  1 (0.8)&0 (0)\\ \hline  
         White&  127 (50.8)& 48 (48)&  59 (48.8)&79 (52.7)\\ \hline  
         More than one race&  36 (14.4)& 14 (14)&  18 (14.9)&22 (14.7)\\ \hline  
         Other&  13 (5.2)& 4 (4)&  4 (3.3)&9 (6.0)\\ \hline  
         Unknown/not reported&  1 (0.4)& 0 (0)&  0 (0)&1 (0.7)\\ \hline  
         \textbf{Ethnicity}&  & &  &\\ \hline  
         Hispanic&  97 (38.8)& 34 (34)&  45 (37.2)&63 (42.0)\\ \hline  
         Not Hispanic&  151 (60.4)& 64 (64)&  74 (61.2)&87 (58.0)\\ \hline  
         Unknown/not reported&  2 (0.8)& 2 (2)&  2 (1.7)&0\\ \hline  
         Average Age (years)&  58. 67$\pm$11.9& 57.74$\pm$ 11.3&  57.59$\pm$ 11.5&59.18$\pm$ 12.1\\ \hline
         Ischemic Strokes& 199 (79.6)& 65 (65)&  65 (64.4)&124 (83.2)\\ \hline 
 Hemorrhagic Strokes& 51 (20.4)& 35 (35)& 36 (35.6)&24 (18.1)\\\hline
 Average time post-stroke upon evaluation (days)& 861$\pm$ 811.2& 891.2$\pm$ 1015.4& 960.6$\pm$ 1114.2& 692$\pm$ 747.88\\\hline 
    \end{tabular}
    \end{center}
    \label{tab:demographics}
\end{table*}

Overall, 60.4\% of the cohort had no weakness throughout the limb. However, among the subjects with some form of weakness, the most common weakness pattern was distal predominant weakness (22.8\% of the overall cohort). Equal weakness throughout the limb represented 12.0\% of the cohort, and proximal predominant weakness represented the smallest subgroup at 3.0\%. We were unable to determine the exact weakness pattern of 1.6\% of the population.

\begin{table*}[!hbt]
    \caption{Distal Predominant Weakness Pattern in Subcortical vs. Non-Subcortical Strokes (significance threshold = 0.05; $\dagger$ Supratentorial p-value = 0.014, $\ddagger$ Whole-Brain p-value = 0.012) and Distal Predominant Weakness Pattern in Ischemic vs. Hemorrhagic Strokes (significance threshold=0.05; p=0.00002)}
    \centering
    \begin{tabular}{|c|c|c|} \hline 
         Stroke Location&  Number -- Supratentorial $\dagger$ & Number -- Whole-Brain $\ddagger$ \\ \hline 
         \textbf{Not Exclusively Subcortical}&\textbf{149}  & \textbf{128}\\ \hline 
         Exhibited Distal Predominant Pattern&  26 (17.4\%)& 23 (17.9\%)\\ \hline 
         Did Not Exhibit Distal Predominant Pattern&123 (82.6\%)& 105 (82.0\%)\\ \hline 
         \textbf{Exclusively Subcortical}&\textbf{101}  & \textbf{122}\\ \hline 
        Exhibited Distal Predominant Pattern & 31 (30.6\%)& 34 (27.9\%)\\ \hline 
         Did not Exhibit Distal Predominant Pattern&70 (69.3\%)& 88 (72.1\%)\\\specialrule{1.5pt}{0pt}{0pt}
Stroke Type &\textbf{Ischemic}&\textbf{Hemorrhagic}\\\hline
 Exhibited Distal Predominant Pattern& 34 (17.1\%)&23 (45.1)\\\hline
 Did Not Exhibit Distal Predominant Pattern& 165 (82.9\%)&28 (54.9)\\\hline
    \end{tabular}
    \label{tab:outcomes}
\end{table*}

\subsection{Primary Analysis - Comparisons between Exclusively Subcortical and Non-Exclusively Subcortical Strokes}

\begin{figure} [!htpb]
    \centering
    \includegraphics[width=0.75\linewidth]{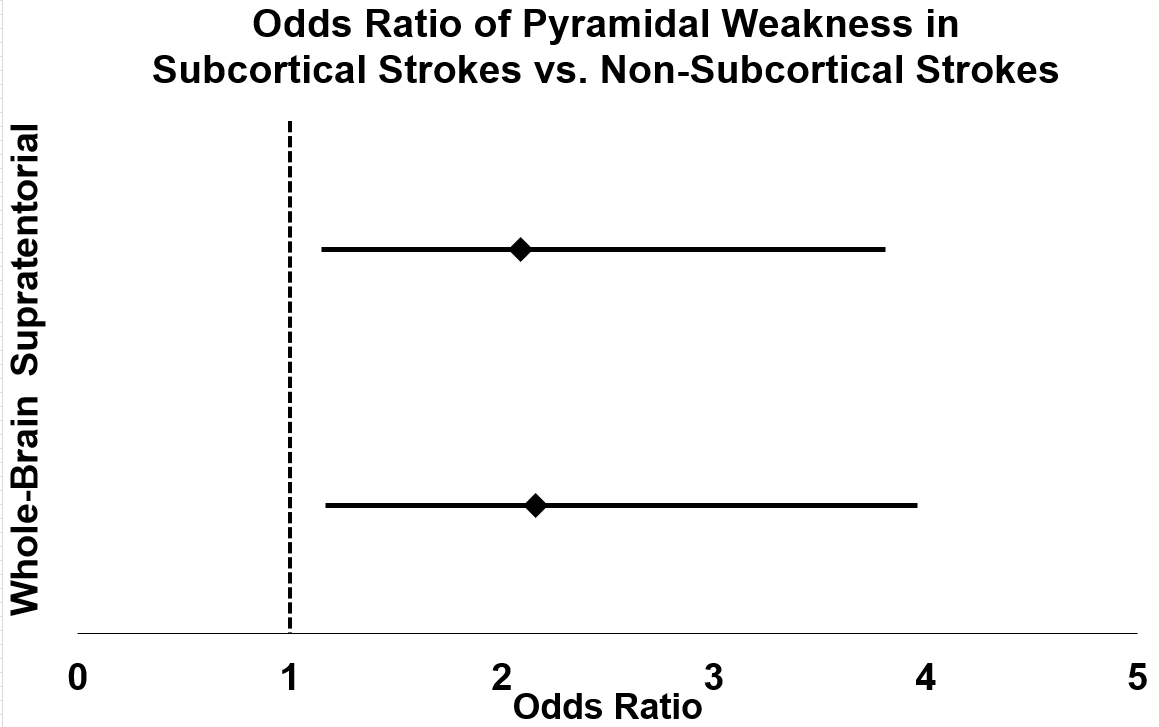}
    \includegraphics[width=0.75\linewidth]{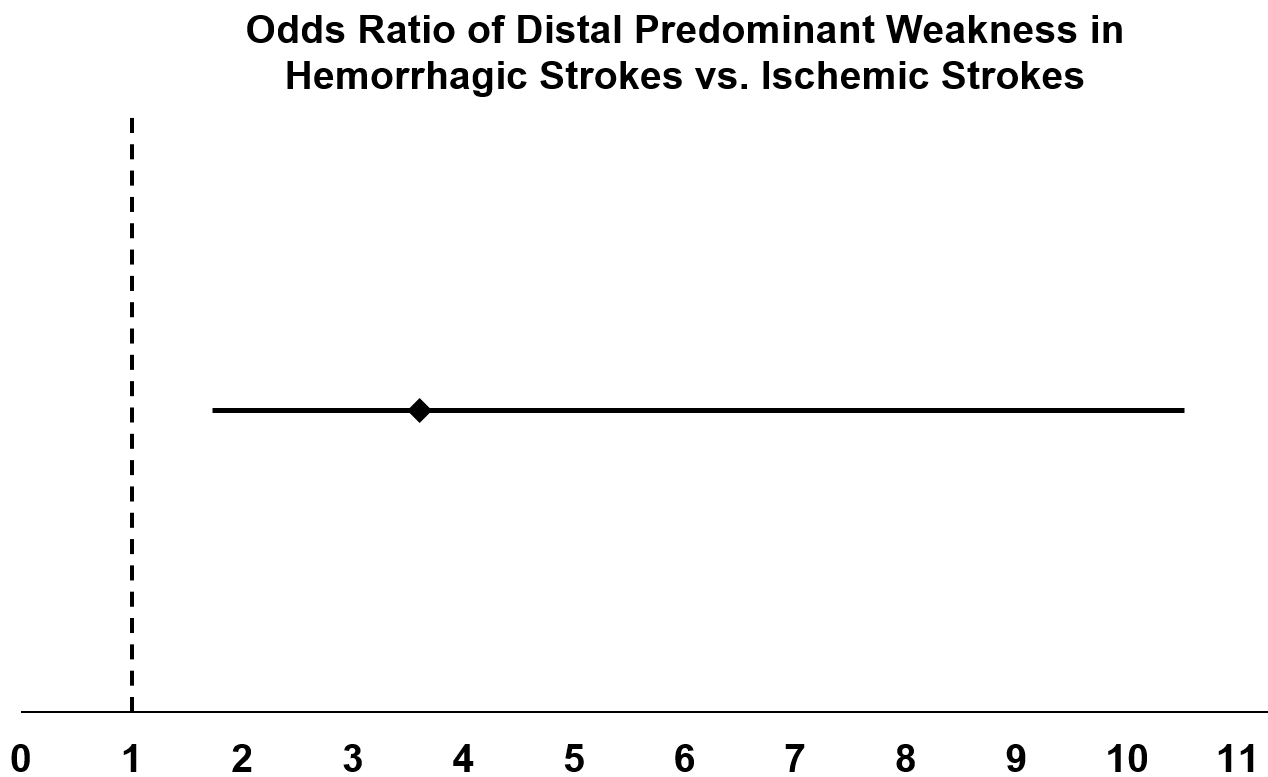}
    \caption{Top: Odds ratio (OR) of distal predominant weakness in exclusively subcortical and non-exclusively subcortical strokes using either supratentorial or whole-brain definition. Bars indicate the 95\% confidence interval. Bottom: OR of distal predominant weakness in hemorrhagic vs. ischemic stroke. Bar represents 95 \% confidence interval.}
    \label{fig:oddsratio}
\end{figure}

First, using the supratentorial definition of subcortical strokes, i.e. the brainstem not being considered subcortical, there were 101 subjects with exclusively subcortical strokes, of whom 31 (30.6\%) exhibited a distal predominant weakness pattern, while 70 did not. Meanwhile, 26 (17.4\%) of the 149 patients with non-exclusively subcortical strokes exhibited this pattern, and 123 did not. This indicates a higher rate of distal predominant weakness in exclusively subcortical strokes (OR = 2.09, 95\% CI: 1.15 - 3.81, p=0.014, $\chi$-squared analysis, Figure \ref{fig:oddsratio}).

Using the whole-brain definition of subcortical strokes, there were 122 exclusively subcortical stroke cases, 34 (27.9\%) of which exhibited a distal predominant weakness pattern, while 72 (72.1 \%) did not. Meanwhile, 128 strokes were considered not exclusively subcortical and 23 (17.9\%) of these exhibited a distal predominant weakness pattern, while 105 did not (Table \ref{tab:outcomes}). Similar to the supratentorial definition, this again indicates a higher prevalence of distal predominant weakness (OR = 2.16, 95\% CI: 1.17 - 3.96, p = 0.012, $\chi$-squared analysis, Figure \ref{fig:oddsratio}).

\begin{figure*}[hbt!]
    \centering
    \includegraphics[width=1\linewidth]{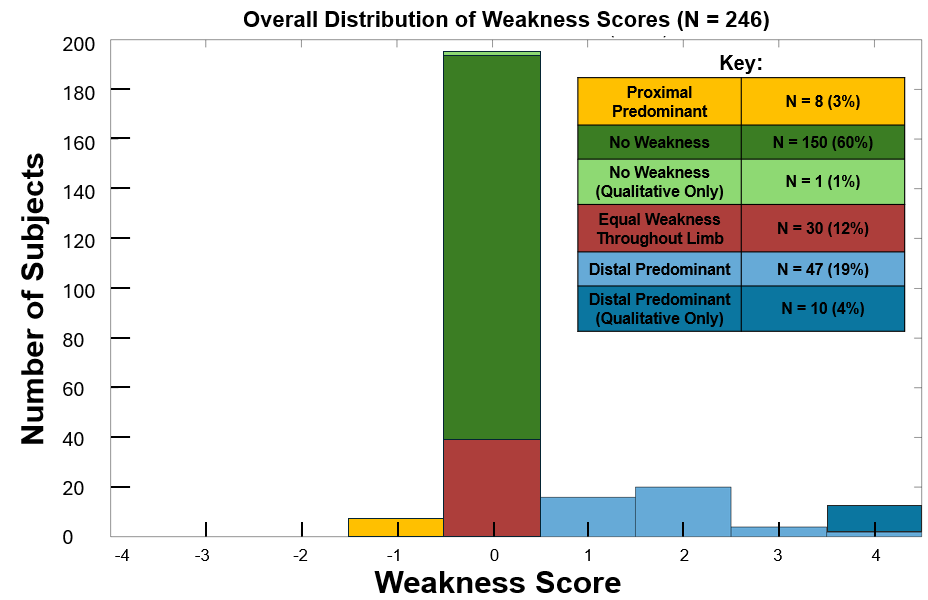}
    \caption{Distribution of the Weakness Score for the entire cohort, including subjects with full strength. 
    Note that four subjects whose qualitative data indicated that they did not exhibit a distal predominant weakness pattern were excluded from this distribution since no further classification could be made.}
    \label{fig:weaknessdistributionscore}
\end{figure*}

An overall distribution of weakness scores for the overall cohort is shown graphically in Figure \ref{fig:weaknessdistributionscore}.

\subsection{Secondary Analyses}

\subsubsection{Localization of Ischemic and Hemorrhagic Strokes}

We found that 66 of 199 (33.1\%) ischemic strokes were exclusively subcortical by the whole-brain definition, while 35 of 51 (68.6\%) hemorrhagic strokes were exclusively subcortical (p = 1$\times$10$^{-5}$, $\chi$-squared test). Meanwhile, 64 of 199 (32.6\%) ischemic strokes were exclusively subcortical by the supratentorial definition. Thirty-five of 51 (68.6\%) hemorrhagic strokes were exclusively subcortical by the supratentorial definition (p = 1$\times$10$^{-5}$, $\chi$-squared test). Of the 199 subjects with ischemic strokes, 34 (17.1\%) exhibited a distal predominant weakness pattern (Table \ref{tab:outcomes}). Twenty-three of the 51 (45.1\%) patients with hemorrhagic stroke exhibited a distal predominant weakness pattern, representing a higher prevalence of distal predominant weakness between (OR = 3.60, 95\% CI: 1.87-6.92, p = 2$\times$10$^{-5}$, $\chi$-square test, Figure \ref{fig:oddsratio}).

\subsubsection{Weakness Scores in Subgroups}

\begin{table*}[hbtp!]
    \centering
    \caption{Frequency of weakness patterns in subgroups. Note that subjects whose weakness scores could not be calculated, due to availability of qualitative data only, were excluded. The cortex and deep structures were selectively reported here as they were the two locations with the most subjects.}
    \label{tab:subgroups}
    \begin{tabular}{|>{\centering\arraybackslash}p{0.18\linewidth}|>{\centering\arraybackslash}p{0.15\linewidth}|>{\centering\arraybackslash}p{0.15\linewidth}|>{\centering\arraybackslash}p{0.15\linewidth}|>{\centering\arraybackslash}p{0.15\linewidth}|}\hline
         Stroke Location&  Proximal Predominant Weakness&  No Weakness&  Equal Weakness Throughout Limb&  Distal Predominant Weakness\\\hline
         Exclusively Subcortial (Supratentorial)&  6 (6\%)&  48 (53\%)&  10 (11\%)&   27 (30\%)\\\hline
         Non-Exclusively Subcortical (Supratentorial)&  2 (1\%)&  101 (70\%)&  20 ( 14\%)&  21 (15\%)\\\hline
         Exclusively Subcortical (Whole-brain)&  6 (5\%)&  61 (55\%)&  14 (13 \%)&  30 (27\%)\\\hline
         Non-Exclusively Subcortical (Whole-brain) &  2 (2\%)&  88 (71 \%)&  16 (13\%)&  18 (15\%)\\\hline
         Cortex&  1 (2\%)&  41 (72\%)&  6 (10\%)&  9 (16\%)\\\hline
 Deep structures& 3 (4\%)& 33 (49\%)& 8 (12\%)&23 (34\%)\\\hline
    \end{tabular}
\end{table*}

Weakness scores were also calculated for multiple selected subgroups, as shown in Table \ref{tab:subgroups}. The cortex and deep structures were selected for reporting in Table Table \ref{tab:subgroups} as they were the two locations with the most subjects.

\section{Discussion}

This study investigated the prevalence of distal predominant weakness patterns in the upper extremities of chronic stroke patients. Our main findings were as follows: (1) chronic stroke subjects with exclusively subcortical strokes were significantly more likely to exhibit the distal predominant weakness pattern than those with non-exclusively subcortical strokes, (2) a distal predominant weakness pattern was the most common pattern of weakness in chronic stroke, regardless of stroke location, (3) a proximal predominant weakness pattern was extremely rare in chronic stroke (4) hemorrhagic strokes were more likely to exhibit a distal predominant weakness pattern. 

The distal predominant weakness pattern was more likely to occur in subjects with exclusively subcortical strokes than non-exclusively subcortical strokes. This observation was present regardless of the supratentorial and whole-brain definitions of exclusively subcortical strokes.
Despite the increased likelihood of distal predominant weakness among the exclusively subcortical population, the majority of subjects were found to have full strength throughout the limb in the chronic phase. 
Furthermore, the distal predominant weakness was the most common weakness pattern among subjects with residual weakness. This finding could be clinically relevant, as physiotherapy and bracing plans can be developed to focus on restoring distal muscle group strength in chronic stroke patients. While proximal muscle groups should not be neglected, the statistical probability that distal muscle groups will be disproportionately affected warrants consideration for extra rehabilitation focus. It also establishes that patients with distal predominant weakness patterns make up a substantial proportion of the chronic stroke population, and therefore further investigation into this phenomenon and rehabilitating these distal segments are warranted.

Although \cite{Beebe:08} did not find an increased likelihood of distal or proximal predominant weakness patterns in chronic post-stroke patients, our findings here are consistent with more recent work from Lin \textit{et al.} \cite{lindj:23}. Specifically, Lin \textit{et al} \cite{lindj:23} examined stroke subjects up to 90 days post ictus and found that a distal predominant weakness pattern was most commonly associated with subcortical lesions. However, only limited comparisons can be drawn between Lin \textit{et al} \cite{lindj:23} and this work due to the measured point in time of recovery. Specifically, Lin \textit{et al.} studied weakness patterns at the subacute stroke time period, whereas this study is in the chronic phase. As such, the differences in findings between these two studies could represent an evolution of patterns over time.
As such, Lin \textit{et al} \cite{lindj:23} found that a significant proportion of subjects ($\sim$ 23\%) had a proximal predominant weakness in the subacute stage, whereas the current study found this pattern to be very rare in the chronic phase (3\%). These observations raise the possibility that stroke patients who exhibit distal predominant weakness are less likely to experience an evolution of weakness pattern, whereas those who start with a proximal predominant weakness in the subacute phase likely either recover their strength or transition to another weakness pattern (i.e., distal predominant). 

Our findings reinforced the established proportions of stroke patients with ischemic and hemorrhagic strokes,

Hemorrhagic strokes were significantly more likely to be located in subcortical regions than ischemic strokes. Conversely, ischemic strokes were much more likely to occur in non-subcortical areas than hemorrhagic strokes. This finding reinforces previous clinical descriptions of hemorrhagic strokes \cite{Kumar:23,An:17,Dastur:17,Lee:12}). Hemorrhagic strokes were more likely to result in a distal predominant weakness pattern than ischemic strokes (45\% vs 17.1\%, respectively). 
This finding is not surprising given that hemorrhagic strokes are more likely to be exclusively subcortical.
Such a difference could inform clinical and rehabilitative decision-making when treating chronic stroke patients. 

The main limitation of this study largely lies with its retrospective nature. Since subjects were not examined by the same provider and through a standardized methodology, this likely introduced variance to how the MMT data were acquired. Additionally, subjects were identified by two different means, as described in greater detail in the methods section. This could introduce a potential sampling bias. The retrospective nature of the study resulted in the inclusion of 15 subjects that only had qualitative data available. Although small (6\% of sample size), this may affect the overall accuracy of the dataset. Finally, the evolution of weakness patterns throughout the time course from ictus to the chronic phase was not directly examined in the study. This was out of the scope of this work's main purpose, which was to establish the prevalence of various weakness patterns in the chronic phase of stroke. Furthermore, the retrospective nature of this study would likely introduce too much variance over multiple time points to provide a meaningful interpretation on weakness pattern evolution.

Future studies should address these limitations by carrying out a prospective examination of distal predominant weakness in a more controlled manner. This would allow for each subject to be uniformly examined to minimize variation in muscle strength testing and to minimize sample bias. This approach could also be utilized to examine the prevalence of weakness patterns in the lower extremities.

\section{Conclusion}

In conclusion, the distal predominant weakness pattern was more prevalent in patients with exclusively subcortical strokes than those with non-exclusively subcortical strokes. 
Furthermore, the distal predominant weakness pattern was the most common weakness pattern seen in chronic stroke. 
The least common pattern found was proximal predominant weakness.
Finally, hemorrhagic strokes were more associated with distal predominant weakness patterns than ischemic strokes. These findings represent a potentially useful clinical guideline and suggest that future studies might be necessary to examine this phenomenon in a more controlled fashion.

\section*{Acknowledgments}
Study funded by University of California, Irvine and NIH R01 \#HD095457

\bibliographystyle{plain}
\bibliography{reference}

\end{document}